\documentclass[useAMS,usenatbib]{mn2e}
\usepackage{url}
\usepackage{stmaryrd}
\usepackage[pdftex]{graphicx}
\usepackage[draft]{hyperref}
\usepackage{amsfonts}

\def\aj{AJ}%
\def\apj{ApJ}%
\def\apjl{ApJ}%
\def\aap{A\&A}%
\def\baas{BAAS}%
\def\icarus{Icarus}%
\def\mnras{MNRAS}%
\def\pasp{PASP}%
\def\skytel{S\&T}%
\def\ssr{Space~Sci.~Rev.}%
\def\nat{Nature}%
\def\iaucirc{IAU~Circ.}%
\def\grl{Geophys.~Res.~Lett.}%
\title[Pluto System Nomenclature]{Proposed Nomenclature for Surface
  Features on Pluto and Its Satellites and Names for Newly Discovered
  Satellites}

\author[E. E. Mamajek et al.]{
Eric E. Mamajek$^{1}$\thanks{E-mail contact: emamajek@pas.rochester.edu},
Valerie A. Rapson$^{2}$,
David A. Cameron$^{1}$,
Manuel Olmedo$^{1,3}$,
\newauthor
Shane Fogerty$^{1}$,
Eric Franklin$^{1}$,
Erini Lambrides$^{4}$,
Imran Hasan$^{5}$,
\newauthor
Richard E. Sarkis$^{1}$,
Stephen Thorndike$^{6}$,
Jason Nordhaus$^{2,7,8}$\\
$^{1}$ Department of Physics \& Astronomy, University of Rochester, 
Rochester, NY, 14627-0171, USA\\
$^{2}$ School of Physics \& Astronomy, Rochester Institute 
of Technology, 54 Lomb Memorial Dr., Rochester, NY, 14623, USA\\
$^{3}$ Instituto Nacional de Astrof\'{i}sica, Optica y Electr\'{o}nica, 
Luis Enrique Erro \#1 C.P. 72840, Tonatzintla, Puebla, M\'{e}xico\\
$^{4}$ American Museum of Natural History, Central Park W \& 79th St., New York, NY 10024\\
$^{5}$ Department of Astronomy, Yale University, P.O. Box 208101, New Haven, CT, 06520-8101, USA\\
$^{6}$ Monroe 2-Orleans BOCES, 3599 Big Ridge Rd., Spencerport, NY, 14559, USA\\
$^{7}$ Center for Computational Relativity and Gravitation, Rochester Institute of Technology, Rochester, NY, 14623, USA\\
$^{8}$ National Technical Institute for the Deaf, Rochester Institute of Technology, Rochester, NY, 14623, USA\\
}
\begin{document}


\pagerange{\pageref{firstpage}--\pageref{lastpage}} \pubyear{2002}

\maketitle

\label{firstpage}

\begin{abstract}
In anticipation of the July 2015 flyby of the Pluto system by NASA's
New Horizons mission, we propose naming conventions and example names
for surface features on Pluto and its satellites (Charon, Nix, Hydra,
Kerberos, Styx) and names for newly discovered satellites.
\end{abstract}

\begin{keywords}
Kuiper Belt objects: individual (Pluto, Charon, Nix, Hydra, Kerberos, Styx) --
planets and satellites: surfaces -- 
solar system: general --
standards
\end{keywords}

\section{Motivation}

Pluto was considered a major planet between its discovery in 1930 by
Clyde Tombaugh \citep{Shapley30, Tombaugh46} and its reclassification
by the International Astronomical Union (IAU) as a dwarf planet in
2006
\citep{vanderHucht08}\footnote{http://www.iau.org/public$\_$press/news/detail/iau0603/}.
Pluto appears to be a remarkably interesting object which sports an
atmosphere, albedo variations, and an extensive satellite system
\citep[e.g.][]{Elliot89, Buie92, Stern92, Owen93, Brown02,
  Pasachoff05, Weaver06, Ward06, Elliot07, Person08, Stern08,
  Lellouch09, Buie10, Merlin10, Tegler10}.
A regularly updated bibliography of studies related to Pluto and its
satellites is maintained by Robert
L. Marcialis\footnote{http://www.lpl.arizona.edu/$\sim$umpire/science/plubib.html}.\\

The NASA New Horizons (NH) Pluto-Kuiper Belt (PKB) mission is
scheduled to fly by the dwarf planet Pluto and its satellite system on
14 July 2015\footnote{http://pluto.jhuapl.edu/}.
Primary mission objectives for NH include investigating the geology,
morphology, and surface composition of Pluto and Charon
\citep{Stern08}. 
High resolution images of Pluto and its satellites will soon be
forthcoming in mid-2015, so discussion of ideas regarding naming
conventions for surface features on these bodies is timely.
The New Horizons team, in coordination with the International
Astronomical Union, has initiated a public campaign for input on
naming of features on Pluto and
Charon\footnote{http://www.ourpluto.org/vote}.
This white paper combines input and discussions from several
astronomers and current and past students in the Rochester area
(mostly currently or recently affiliated with University of Rochester
[AST 111 \& 142 classes] and Rochester Institute of Technology).

\section{Nomenclature}

Historically, many solar system objects have been named for
mythological deities, as well as people and places from myths and
classic literature from various
cultures\footnote{http://planetarynames.wr.usgs.gov/}.
The general rules and conventions of planetary nomenclature have been
outlined by the
IAU\footnote{http://planetarynames.wr.usgs.gov/Page/Rules}.
The IAU Working Groups for Planetary System Nomenclature (WGPSN)
maintains a compilation of categories of surface features on solar
system bodies along with the naming convention for each
category\footnote{http://planetarynames.wr.usgs.gov/Page/Categories}.
The Working Group maintains a list of reputable sources which contain
the spellings and descriptions of people, places, and things, which
have been used for naming planetary
features\footnote{http://planetarynames.wr.usgs.gov/References}
\citep[e.g.][]{Guirand77}.\\

Thus far, only two naming conventions have been used for the Pluto
system. Surface features of Pluto are to be named for {\it
  ``Underworld
  deities''} \footnote{http://planetarynames.wr.usgs.gov/Page/Categories}.
The IAU WGPSN and SBN \footnote{http://www.ss.astro.umd.edu/IAU/csbn/}
have adopted the following for the naming of Pluto's satellites: {\it
  ``Satellites in the plutonian system are named for characters and
  creatures in the myths surrounding Pluto (Greek Hades) and the
  classical Greek and Roman
  Underworld.''}\footnote{http://planetarynames.wr.usgs.gov/Page/Planets}.\\

The dwarf planet Pluto and its largest satellite Charon manifest
significant albedo variations, likely reflecting different types of
complex terrain.
Neptune's largest satellite Triton, which may be the most Pluto-like
body yet imaged by spacecraft, has its surface features categorized by
12 different classes: {\it catenae, cavi, craters, dorsa, fossae,
  maculae, paterae, planitiae, plana, plumes, regiones, sulci}.
There is some spectroscopic evidence (based on the presence of
crystalline water ice and ammonia hydrates) that Charon may be
experiencing cryovolcanism \citep{Cook07}.
It seems likely that imagery of the surfaces of Pluto and Charon may
warrant a number of toponymic classes similar to that of Triton.
In what follows, we summarize some suggestions which expand upon the
existing IAU naming themes for the Pluto system.\\


\subsection{Pluto}

Pluto was discovered in 1930 by Clyde Tombaugh \citep{Shapley30,
  Tombaugh46, Tombaugh60, Tombaugh97}. In Greek mythology, Pluto was
ruler of the underworld Hades, and represented a deity of wealth 
and treasure \citep{Guirand77}.\\

\begin{itemize}

\item {\it Albedo features, Planitiae, Plana, Terrae, Dorsa, Maculae,
  Mensae, Tesserae:} Deceased people and places associated with the
  discovery and characterization of Pluto:
Tombaugh \citep[Clyde William, 1906-1997;][]{Tombaugh46, Tombaugh60,
  Tombaugh97},
Lowell (Percival Lawrence, 1855-1916; began effort which lead to
discovery of Pluto),
Burney (Venetia Katharine Douglas Burney, 1918-2009; suggested name
for Pluto),
Kuiper \citep[Gerald Peter, 1905-1973;][]{Kuiper50,Kuiper57},  
Elliot \citep[James Ludlow, 1943-2011;][]{Elliot89,Elliot92,Elliot03,Elliot07}\footnote{http://web.mit.edu/newsoffice/2011/obit-elliot.html}, 
Rabe \citep[Eugene Karl, 1911-1974;][]{Rabe57,Rabe58},
Hunten \citep[Donald M., 1925-2010;][]{Hunten82}\footnote{http://aas.org/obituaries/donald-m-hunten-1925-2010}, 
Simonelli \citep[Damon Paul, 1959-2004;][]{Simonelli89, Buratti05},
Bower \citep[Ernest Clare, 1890-1964;][]{Bower30, Bower31, Bower34,
  Hockey09},
Whipple \citep[Fred Lawrence, 1906-2004;][]{Bower30,Yeomans04},
Hardie \citep[Robert, 1923-1989;][]{Walker55, Hardie65, Tenn07},
Flagstaff, 
Coconino, 
Arizona (locations of Lowell observatory and USNO
Flagstaff station where Pluto and Charon were discovered,
respectively).\\

\item {\it Craters:} Underworld deities and locations from mythologies
  around the world, excluding psychopomps (reserved for Charon; see
  \S2.2).  Examples: Mictlan (the Aztec underworld), Wepwawet
  \citep[ancient Egyptian mythology;][]{Hart90}, etc.\\

\item {\it Catenae, Cavi, Chasmata, Fossae, Labryinthi, Montes,
  Paterae, Rupes, Scopuli, Sulci, Valles, Tholi:} Words for
  "cold" in extinct or endangered
  languages\footnote{http://www.unesco.org/culture/languages-atlas/index.php} \footnote{http://www.endangeredlanguages.com/}.
  This is analagous to use of terms for ``hot'' for {\it montes} on
  Mercury\footnote{http://planetarynames.wr.usgs.gov/Page/Categories}.
  Words from documented extinct or endangered languages provide a
  nearly limitless (and thus far surprisingly underutilized) reservoir
  of names that may be used for celestial nomenclature.
Examples: 
Nirum \citep[Yaygir language, Australia;][]{Crowley79}, 
Hanglu \citep[Siraya language, Taiwan;][]{Adelaar11}, 
Julu \citep[Present-day Great Andamanese, Andaman Islands;][]{Abbi13},
etc.\\

Other potential themes for Pluto features: Geological features and
archaeological sites of northern Arizona (region where both Pluto and
Charon were discovered), names of famous coins or mints (given Pluto's
association with wealth).\\

\end{itemize}


\subsection{Charon (134340 Pluto I)}

Charon was discovered by \citet{Christy78} and ``{\it named after the
  Greek mythological boatman who ferried souls across the river Styx
  to Pluto for
  judgement}''\footnote{http://planetarynames.wr.usgs.gov/Page/Planets}.\\

\begin{itemize}

\item {\it Albedo features, Planitiae, Plana, Terrae:} People
  associated with the discovery and characterization of Pluto's
  satellites, and mythological psychopomps (deities responsible for
  guiding spirits to the afterlife).  Charon was discovered so
  recently (1978), that most notable scientists associated with its
  study are still alive, with a notable exception being Harrington
  \citep[Robert Sutton, 1942-1993;][]{Christy78} who co-discovered
  Charon with Christy and was the first to calculate a dynamical mass
  for the Pluto-Charon
  system\footnote{http://ad.usno.navy.mil/wds/history/harrington.html}.
  Mythological counterparts to Charon abound in the literature, with
  examples: Anubis \citep[ancient Egyptian;][]{Hart90}, Muut
  \citep[Cahuilla people of southern California;][]{Bean74}, Namtar
  \citep[Mesopotamian;][]{Black04}, Ixtab (Mayan), etc.\\

\item {\it Craters:} Names of characters, places, and starships
  associated with the Star Trek
  series of television shows and movies by American screenwriter and
  producer Gene Roddenberry (1921-1991).  Examples: Kirk, Spock,
  McCoy, Sulu, Uhura, Chekov, Scotty, etc. \citep{Okuda99}.\\

  Gene Roddenberry worked on Star Trek between the mid-1960s and
  approximately 1990, which bracketed the discovery epoch of Charon
  (1978), the epoch of the Pluto-Charon mutual events (eclipses) which
  constrained the sizes of Pluto and Charon (1985-1990). It should be
  noted that there was popular support for naming one of the newly
  discovered satellites P4 or P5 ``Vulcan''. While ``Vulcan'' is
  obviously linked to a deity in classical mythology, its popularity
  was largely based on its use as the name of a fictional planet in
  Star
  Trek\footnote{http://www.space.com/21814-pluto-moons-named-kerberos-styx.html}. There
  is precedent of naming surface features on planetary satellites from
  science fiction and fantasy works from the 20th century. On Titan,
  the IAU has adopted the names of characters of Middle Earth from the
  novels of J.R.R. Tolkien for {\it colles} features, mountains and
  peaks from Middle Earth for {\it montes} peaks, and the names of
  planets from the Dune novels by Frank Herbert
  \citep[e.g.][]{Herbert65} for {\it planitiae} and {\it
    labyrinthi}. There are also asteroids named for fictional
  characters or shows from recent decades: e.g. (9007) James Bond and
  (13681) Monty Python being recent examples. \\

\item {\it Dorsa, Maculae, Mensae, Montes, Paterae, Tesserae, Tholi:}
  Notable science fiction authors\footnote{The names and birth/death
    dates listed for the authors are as listed in Wikipedia
    (http://en.wikipedia.org), hence further verification is clearly
    warranted. A website listing science fiction works that involve
    Pluto as a setting is compiled by Steven H. Silver:
    http://www.sfsite.com/~silverag/pluto.html.} who have written
  stories where Pluto and/or its satellites are featured.  Examples:
Coblentz \citep[Stanton Arthur, 1896-1982;][]{Coblentz31, Coblentz34}, 
Gallun \citep[Raymond Zinke, 1911-1994;][]{Gallun35}, 
Heinlein \citep[Robert Anson, 1907-1988;][]{Heinlein58}, 
Lovecraft \citep[Howard Phillips, 1890-1937;][]{Lovecraft30},
Simak \citep[Clifford Donald, 1904-1988;][]{Simak73},
Starzl \citep[Roman Frederick, 1899-1976;][]{Starzl31a, Starzl31b},
Walters \citep[Hugh, 1910-1993;][]{Walters73},
Weinbaum \citep[Stanley Grauman; 1902-1935; ][]{Weinbaum35}, 
Williamson \citep[Jack, 1908-2006;][]{Williamson33, Williamson50}.
Kornbluth (Cyril M., 1923-1958) and Pohl (Frederik George, Jr.,
1919-2013) coauthored three science fiction stories with Pluto as a
setting under the pseudonyms S.D. Gottesman and Dennis Lavond
\citep{Gottesman40, Gottesman41, Lavond41}. \\

The list of science fiction authors is clearly heavy in American and
western authors, so a broader survey of the international science
fiction literature for notable deceased authors is obviously warranted
if this nomenclature theme is adopted.  Deceased Slavic authors who
have written stories including Pluto as a setting are also not
difficult to find: Shalimov (Alexander), Lukodjanov, Bulychov (Kir,
alias of Igor Vsevolodovich Mozheiko), Bil\'{e}nkin (Dmitri
Aleks\'{a}ndrovitch; 1933-1987), Yemfrimov (Ivan Antonovich,
1908-1972)\footnote{Many thanks to Valentin Ivanov (ESO) for pointing
  us to these science fiction authors.}.  Unfortunately, at present we
have only found incomplete translated bibliographic information on
their works and birth/death dates.\\






\item {\it Catenae, Cavi, Chasmata, Fossae, Labyrinthi, Rupes,
  Scopuli, Sulci, Valles:} Small towns, settlements, or
  islands serviced predominantly by ferry. Examples:
Nias (Indonesia),
Fogo (Canada),
Robben (South Africa), etc.

\end{itemize}


\subsection{Nix (134340 Pluto II = S/2005 P 2)}

Nix was discovered in 2005 by \citet{Weaver05}, and the IAU WGPSN
approved its designation in 2006.  The adopted IAU spelling of the
name is the Egyptian spelling of the Greek {\it
  Nyx}\footnote{http://planetarynames.wr.usgs.gov/Page/Planets}. In
Hesiod's {\it Theogeny} \citep{Hesiod14}, Nyx was one of the first
offspring of Chaos, ``{\it goddess of darkness and night, mother of
  Charon.}''\\

\begin{itemize}

\item {\it Surface Features}: Offspring of the Greek god Nyx.  Nyx and
  her family are detailed in \citet{Hesiod14}\footnote{Nyx's place in
    the family tree of Greek mythological deities is shown at
    http://www.theoi.com/TreeHesiod.html}.  Examples: Moros, Ker,
  Hypnos, Momos, Oizys, etc. \citep{Hesiod14}.

\end{itemize}


\subsection{Hydra (134340 Pluto III = S/2005 P 1)}

Hydra was discovered in 2005 by \citet{Weaver05}, and the IAU WGPSN
approved its desigation in 2006. The Lernaen Hydra ``{\it born of
  Typhon and Echidna, was an enormous serpent with nine heads,}''
slain by Hercules \citep{Guirand77}.\\

\begin{itemize}

\item {\it Surface features:} People and places associated with the
  Greek mythological water monster Hydra. Examples: Typhon, Echidna,
  Heracles or Hercules, Lerna, Iolaus, etc. \citep{Guirand77}.\\

Note that ``{\it Sea creatures from myth and
  literature}''\footnote{http://planetarynames.wr.usgs.gov/Page/Categories}
has already been adopted by the IAU as the naming theme for {\it
  maria} on Titan, and so should not be reused for Hydra.

\end{itemize}


\subsection{Kerberos (134340 Pluto IV = S/2011 (134340) 1)} 

Kerberos was discovered by \citet{Showalter11}.  While the name
Cerberus was proposed, it was already in use for an asteroid, and the
name of the Greek counterpart Kerberos was adopted by the
IAU. Kerberos was ``{\it the hound of Hades}'' \citep{Hesiod14}, a
``{\it monstrous watch-dog with fifty heads and a voice of bronze}'',
although he was usually depicted with only three heads
\citep{Guirand77}.\\

\begin{itemize}

\item {\it Surface Features:} Canine deities and monsters from
  mythology around the world.  Chimera (Greek), Garmr or Garm (Norse),
  Fenrir \citep[Norse;][]{Guirand77}, Amarok (Inuit), Asena (Turkic),
  Kishi (Angola), C\'{u}-S\'{i}th (Scotland), Inugami (Japan).\\

  Thus far, no naming theme for any solar system body focuses on
  canine deities or monsters.\\

\end{itemize}


\subsection{Styx (134340 Pluto V = S/2012 (134340) 1)}

Styx was discovered by \citet{Showalter12}. Styx is ``{\it the
  infernal river}'' that separates Earth (the land of the living) from
the underworld (the land of the dead) in Greek mythology. Styx was
personified as a nymph born of Tethys and Oceanus \citep{Guirand77}.

\begin{itemize}

\item {\it Surface Features:} Rivers of mythological underworlds.
  Examples from Greek mythology: Acheron, Cocytus, Lethe,
  Phlegethon \citep{Guirand77}.\\

If a wider variety of nomenclature is needed, the nymph Styx from
Greek mythology also had offspring with Pallas: Bia, Kratos, Nike, and
Zelos \citep{Guirand77}.\\

\end{itemize}


\section{New Satellites}

Pluto's known satellites have been named Charon, Nix, Hydra, Kerberos,
and Styx after mythological characters associated with the underworld
from Roman or Greek mythology. Current models for forming Pluto and
Charon favor a giant impact origin for the satellite system, where the
outer satellites formed from an ice-rich, post-collision debris disk
\citep[][]{Canup05,Stern06, Canup11,Kenyon14,Bromley15}.  Recent
models by \citet{Kenyon14} have predicted that the Pluto system may
contain additional small satellites ($<$few km) beyond the orbit of
Hydra at orbital radii of $\sim$70-200 R$_{Pluto}$, with small
inclinations with respect to Pluto-Charon. This region may contain an
extremely low optical depth debris ring as well \citep{Kenyon14}. It
is also possible that the small satellites Nix and Hydra could harbor
tiny coorbital satellites \citep{PiresDosSantos11}.\\

Based on input from the public, Mark Showalter submitted the names
Vulcan and Cerberus to the IAU WGPSN and SBN committees for the
satellites dubbed P4 and P5 (now Kerberos and Styx).  Vulcan was
rejected on the grounds that the name had been widely used to refer to
a hypothetical planet that may exist closer to the Sun than Mercury
(ruled out), and we are now left with the term 'vulcanoids' attached
to the hypothetical population of asteroids which may orbit the Sun
closer than Mercury's orbit (not ruled out). Cerberus was rejected due
to its previous use with the asteroid 1865 Cerberus, however a
transliteration of the Greek spelling {\it Kerberos} was
adopted\footnote{http://www.iau.org/public$\_$press/news/detail/iau1303/}.\\

Showalter received several other good candidate names, including
Elysium, Tartarus, Tantalus, Sisyphus, Orthrus, Melinoe, Hecate,
Thanatos. These are briefly discussed and described on Showalter's
blog\footnote{http://cosmicdiary.org/mshowalter/2013/02/14/opening-up-the-gates-of-hell/}.
A search of the IAU Minor Planet Center compilation of minor planet
names\footnote{http://www.minorplanetcenter.net/iau/lists/MPNames.html}
shows that as of 8 March 2015, there are asteroids named (1866)
Sisyphus, (2102) Tantalus, and (100) Hekate (close enough to Hecate
that 'Hecate' should probably not be considered). However, there are
no asteroids named {\it Tartarus}, {\it Orthrus}, {\it Melinoe}, or
{\it Thanatos}. Arguably, these are all excellent candidate satellite
names.\\


{\it\bf Tartarus} (Latin) or {\it\bf Tartaros} (Greek) represents the
lowest region of the underworld -- the region below even Hades where
the Titans were imprisoned \citep{Autenrieth1891, Guirand77}.\\


{\it\bf Orthrus} or {\it\bf Orthus} was a dog with two heads and the
tail of a serpent from Greek mythology, and sibling of Kerberos
(Cerberus). Orthrus, along with his master Eurytion, guarded a herd of
red oxen controlled by the monster Geryon. Orthrus (along with
seemingly every other interesting beast in Greek mythology) was slain
by Hercules. The name was spelled as {\it Orthus} in Hugh
Evelyn-White's English translation of Hesiod's {\it Theogeny}
\citep{Hesiod14}, and referred to as ``{\it Orthus, the hound of
  Geryones, born of Echidna and Typhaon, slain by Heracles}''.  The
same monster is spelled as {\it Orthrus} in \citet{Guirand77}. There
is, however, an asteroid (2329) Orthos, of similar spelling.\\


{\it\bf Melinoe} was an underworld goddess in Greek mythology who made
nightly earth visits, spreading fear with her ghastly companions. Her
name translates rouhgly to ``dark mind'', and she is described in
detail in Orphic Hymn LXX ``{\it To Melinoe, the Fumigation from
  Aromatics}'' \citep{Taylor1896, Athanassakis13}. Representing her
mixed heritage as the daughter of Zeus and Persephone, Melinoe's limbs
were dark on one side, but white on the other \citep{Taylor1896}.
Melinoe may be a particularly appropriate name for a new satellites
with high contrast surface features.\\


{\it\bf Thanatos} (Greek) or {\it\bf Mors} (Latin) was the god of
non-violent
death\footnote{http://www.theoi.com/Daimon/Thanatos.html}. He is often
depicted with his twin brother, Hypnos, who was the god of
sleep. Thanatos and Hypnos were sons of Nyx (goddess of night) and
Erebos (god of darkness).  Thanatos could rarely be defeated, tricked,
or captured by Greek gods and heroes (Heracles, Sisyphus) who wished
to prevent a death or escape their own death
\citep{Hesiod14}. Thanatos was also the subject of an Orphyic hymn
``{\it To Thanatos, Fumigation from Manna}'' \citep{Taylor1896,
  Hansen04, Athanassakis13}. With Thanatos being a son of Nyx, its
name could be used for a surface feature on Nyx instead.\\



\begin{thebibliography}{99}

\bibitem[\protect\citeauthoryear{Abbi}{2013}]{Abbi13} Abbi, A., 2013,
  {\it A Grammar of the Great Andamanese Language: An Ethnolinguistic
    Study}, Brill, Leiden

\bibitem[\protect\citeauthoryear{Adelaar}{2011}]{Adelaar11} Adelaar,
  A.~K., 2011, {\it Siraya: Retrieving the Phonology, Grammer and
    Lexicon of a Dormant Formosan Language}, Walter de Gruyter GmbH \&
  Co. GK, Berlin

\bibitem[\protect\citeauthoryear{Athanassakis}{2013}]{Athanassakis13}
  Athanassakis, A., 2013, {\it The Orphic hymns: Text, Translation,
    and Notes}, Johns Hopkins University Press, Baltimore

\bibitem[\protect\citeauthoryear{Autenrieth}{1891}]{Autenrieth1891}
  Autenrieth, G., 1891, {\it A Homeric Dictionary for Schools and
    Colleges}, Harper and Brothers, New York

\bibitem[\protect\citeauthoryear{Andersson}{1978}]{Andersson78}
  Andersson, L.~E.\ 1978, \baas, 10, 586

\bibitem[\protect\citeauthoryear{Bean}{1974}]{Bean74} Bean, L.~J.,
  1974, {\it Mukat's People: The Cahuilla Indians of Southern
    California}, University of California Press, Berkeley

\bibitem[\protect\citeauthoryear{Black et al.}{2004}]{Black04} Black,
  J., Green, A., \& Rickards, T., 2004, {\it Gods, Demons and Symbols
    of Ancient Mesopotamia: An Illustrated Dictionary}, 2nd edition
  reprint, The British Museum Press, London

\bibitem[\protect\citeauthoryear{Bower}{1931}]{Bower31} Bower,
  E.~C.\ 1931, Ph.D.~Thesis, University of California, Berkeley

\bibitem[\protect\citeauthoryear{Bower}{1934}]{Bower34} Bower,
  E.~C.\ 1934, Lick Observatory Bulletin, 17, 53

\bibitem[\protect\citeauthoryear{Bower \& Whipple}{1930}]{Bower30}
  Bower, E.~C., \& Whipple, F.~L.\ 1930, \pasp, 42, 236

\bibitem[\protect\citeauthoryear{Bower et al.}{1933}]{Bower33} Bower,
  E.~C., Anderson, C.~M., \& Wyse, A.~B.\ 1933, Lick Observatory
  Bulletin, 16, 115

\bibitem[\protect\citeauthoryear{Bromley \& Kenyon}{2015}]{Bromley15}
  Bromley, B.~C., \& Kenyon, S.~J.\ 2015, submitted to ApJ,
  arXiv:1503.06805

\bibitem[\protect\citeauthoryear{Buie et al.}{1992}]{Buie92} Buie,
  M.~W., Tholen, D.~J., \& Horne, K.\ 1992, \icarus, 97, 211

\bibitem[\protect\citeauthoryear{Buie et al.}{2010}]{Buie10} Buie,
  M.~W., Grundy, W.~M., Young, E.~F., Young, L.~A., \& Stern,
  S.~A.\ 2010, \aj, 139, 1128

\bibitem[\protect\citeauthoryear{Buratti \& Veverka}{2005}]{Buratti05}
  Buratti, B.~J., \& Veverka, J.\ 2005, \baas, 37, 1557

\bibitem[\protect\citeauthoryear{Brown}{2002}]{Brown02} Brown,
  M.~E.\ 2002, Annual Review of Earth and Planetary Sciences, 30, 307

\bibitem[\protect\citeauthoryear{Brown et al.}{2005}]{Brown05} Brown,
  M.~E., Trujillo, C.~A., \& Rabinowitz, D.~L.\ 2005, \apjl, 635, L97

\bibitem[\protect\citeauthoryear{Canup}{2005}]{Canup05} Canup,
  R.~M.\ 2005, Science, 307, 546

\bibitem[\protect\citeauthoryear{Canup}{2011}]{Canup11} Canup,
  R.~M.\ 2011, \aj, 141, 35

\bibitem[\protect\citeauthoryear{Cheng et al.}{2008}]{Cheng08} Cheng,
  A.~F., Weaver, H.~A., Conard, S.~J., et al.\ 2008, \ssr, 140, 189

\bibitem[\protect\citeauthoryear{Christy \&
    Harrington}{1978}]{Christy78} Christy, J.~W., \& Harrington,
  R.~S.\ 1978, \aj, 83, 1005

\bibitem[\protect\citeauthoryear{Coblentz}{1931}]{Coblentz31}
  Coblentz, S.\ 1931, ``Into Plutonian Depths.'' {\it Wonder Stories
    Quarterly}, Spring 1931

\bibitem[\protect\citeauthoryear{Coblentz}{1934}]{Coblentz34}
  Coblentz, S.\ 1934, ``Riches for Pluto.'' {\it Astounding}, December
  1934

\bibitem[\protect\citeauthoryear{Cook et al.}{2007}]{Cook07} Cook,
  J.~C., Desch, S.~J., Roush, T.~L., Trujillo, C.~A., \& Geballe,
  T.~R.\ 2007, \apj, 663, 1406

\bibitem[\protect\citeauthoryear{Crowley}{1979}]{Crowley79} Crowley,
  T., 1979, {\it Handbook of Australian Languages, Volume 1}, Johns
  Benjamins Publishing, eds. R.~M.~W Dixon \& B.~J. Blake

\bibitem[\protect\citeauthoryear{Elliot et al.}{1989}]{Elliot89}
  Elliot, J.~L., Dunham, E.~W., Bosh, A.~S., et al.\ 1989, \icarus,
  77, 148

\bibitem[\protect\citeauthoryear{Elliot \& Young}{1992}]{Elliot92}
  Elliot, J.~L., \& Young, L.~A.\ 1992, \aj, 103, 991

\bibitem[\protect\citeauthoryear{Elliot et al.}{2003}]{Elliot03}
  Elliot, J.~L., Ates, A., Babcock, B.~A., et al.\ 2003, Nature, 424,
  165

\bibitem[\protect\citeauthoryear{Elliot et al.}{2007}]{Elliot07}
  Elliot, J.~L., Person, M.~J., Gulbis, A.~A.~S., et al.\ 2007, \aj,
  134, 1

\bibitem[\protect\citeauthoryear{Gallun}{1935}]{Gallun35} Gallun,
  R.~Z.\ 1935, ``Blue Haze on Pluto.'' {\it Astounding}, June 1935

\bibitem[\protect\citeauthoryear{Gottesman}{1940}]{Gottesman40}
  Gottesman, S.~D.\ 1940, ``King Cole of Pluto.'' {\it Super Science
    Stories}, May 1940

\bibitem[\protect\citeauthoryear{Gottesman}{1941}]{Gottesman41}
  Gottesman, S.~D.\ 1941, ``The Castle on Outerplanet.'' {\it Uncanny
    Tales}, October 1941


\bibitem[\protect\citeauthoryear{Guirand}{1977}]{Guirand77} Guirand,
  F., 1977, {\it New Larousse Encyclopedia of Mythology}, translated
  by R. Aldington \& D. Ames and revised by a panel of editorial
  advisers from the {\it Larousse Mythologie G\'{e}n\'{e}rale}, ed.
  F. Guirand. New edition 1968, thirteenth impression 1977. The Hamlyn
  Publishing Group Limited, New York

\bibitem[\protect\citeauthoryear{Green}{2006}]{Green06} Green,
  D.~W.~E.\ 2006, \iaucirc, 8723, 1

\bibitem[\protect\citeauthoryear{Hansen}{2004}]{Hansen04} Hansen, W.,
  2004, {\it Handbook of Classical Mythology}, ABC-CLIO, Santa
  Barbara, CA (ISBN 1-57607-226-6)

\bibitem[\protect\citeauthoryear{Hardie}{1965}]{Hardie65} Hardie,
  R.\ 1965, \aj, 70, 140

\bibitem[\protect\citeauthoryear{Hart}{1990}]{Hart90} Hart, G., 1990,
  {\it Egyptian Myths: Legendary Past Series}. University of Texas
  Press

\bibitem[\protect\citeauthoryear{Heinlein}{1958}]{Heinlein58}
  Heinlein, R.~A., 1958, {\it Have Spacesuit, Will Travel}. Scribner

\bibitem[\protect\citeauthoryear{Herbert}{1965}]{Herbert65} Herbert,
  F., 1965, {\it Dune}, Chilton Books

\bibitem[\protect\citeauthoryear{Hesiod}{1914}]{Hesiod14} Hesiod,
  1914, {\it The Homeric Hymns and Homerica with an English
    Translation by Hugh G. Evelyn-White}. Theogony. Harvard University
  Press, Cambridge MA.

\bibitem[\protect\citeauthoryear{Hockey et al.}{2009}]{Hockey09}
  Hockey, T., Trimble, V., Williams, T.~R., et al.\ 2009, The
  Biographical Encyclopedia of Astronomers, Edited by Thomas Hockey,
  Virginia Trimble, Thomas R.~Williams, Katherine Bracher, Richard
  A.~Jarrell, Jordan D.~March{\'e}, and F.~Jamil Ragep.~Berlin,
  Springer: 2009, ISBN: 978-0-387-35133-9.

\bibitem[\protect\citeauthoryear{Hunten \& Watson}{1982}]{Hunten82}
  Hunten, D.~M., \& Watson, A.~J.\ 1982, \icarus, 51, 665

\bibitem[\protect\citeauthoryear{Kenyon \& Bromley}{2014}]{Kenyon14}
  Kenyon, S.~J., \& Bromley, B.~C.\ 2014, \aj, 147, 8

\bibitem[\protect\citeauthoryear{Kuiper}{1950}]{Kuiper50} Kuiper,
  G.~P.\ 1950, Pub. Ast. Soc. Pacific, 62, 133

\bibitem[\protect\citeauthoryear{Kuiper}{1957}]{Kuiper57} Kuiper,
  G.~P.\ 1957, \apj, 125, 287

\bibitem[\protect\citeauthoryear{Lavond}{1941}]{Lavond41} Lavond,
  D.\ 1941, ``A Prince of Pluto.'' {\it Future Fiction}, April 1941

\bibitem[\protect\citeauthoryear{Lellouch et al.}{2009}]{Lellouch09}
  Lellouch, E., Sicardy, B., de Bergh, C., et al.\ 2009, \aap, 495,
  L17

\bibitem[\protect\citeauthoryear{Lovecraft}{1930}]{Lovecraft30}
  Lovecraft, H.~P., 1931, ``The Whisperer in Darkness''. {\it The
    Dunwich Horror and Others}.

\bibitem[\protect\citeauthoryear{Merlin et al.}{2010}]{Merlin10}
  Merlin, F., Barucci, M.~A., de Bergh, C., et al.\ 2010, \icarus,
  210, 930

\bibitem[\protect\citeauthoryear{Okuda et al.}{1999}]{Okuda99} Okuda,
  M., Okuda, D., \& Mirek, D., 1999, {\it The Star Trek Encyclopedia:
    A Reference Guide to the Future}. Pocket Books: 1999, ISBN:
  978-0671536091

\bibitem[\protect\citeauthoryear{Olkin et al.}{2007}]{Olkin07} Olkin,
  C.~B., Young, E.~F., Young, L.~A., et al.\ 2007, \aj, 133, 420

\bibitem[\protect\citeauthoryear{Owen et al.}{1993}]{Owen93} Owen,
  T.~C., Roush, T.~L., Cruikshank, D.~P., et al.\ 1993, Science, 261,
  745

\bibitem[\protect\citeauthoryear{Pasachoff et al.}{2005}]{Pasachoff05}
  Pasachoff, J.~M., Souza, S.~P., Babcock, B.~A., et al.\ 2005, \aj,
  129, 1718

\bibitem[\protect\citeauthoryear{Person et al.}{2008}]{Person08}
  Person, M.~J., Elliot, J.~L., Gulbis, A.~A.~S., et al.\ 2008, \aj,
  136, 1510

\bibitem[\protect\citeauthoryear{Pires Dos Santos et
    al.}{2011}]{PiresDosSantos11} Pires Dos Santos, P.~M., Giuliatti
  Winter, S.~M., \& Sfair, R.\ 2011, \mnras, 410, 273

\bibitem[\protect\citeauthoryear{Rabe}{1957a}]{Rabe57} Rabe, E.\ 1957, \apj, 125, 290 

\bibitem[\protect\citeauthoryear{Rabe}{1957b}]{Rabe58} Rabe, E.\ 1957, \apj, 126, 240

\bibitem[\protect\citeauthoryear{Shapley}{1930}]{Shapley30} Shapley, H.\ 1930, \iaucirc, 255, 1

\bibitem[\protect\citeauthoryear{Showalter et al.}{2011}]{Showalter11}
  Showalter, M.~R., Hamilton, D.~P., Stern, S.~A., et al.\ 2011,
  \iaucirc, 9221, 1

\bibitem[\protect\citeauthoryear{Showalter et al.}{2012}]{Showalter12}
  Showalter, M.~R., Weaver, H.~A., Stern, S.~A., et al.\ 2012,
  \iaucirc, 9253, 1

\bibitem[\protect\citeauthoryear{Simak}{1973}]{Simak73} Simak,
  C.~D.\ 1973, ``Construction Shack.'' {\it Worlds of If}, Jan/Feb
  1973

\bibitem[\protect\citeauthoryear{Simonelli \&
    Reynolds}{1989}]{Simonelli89} Simonelli, D.~P., \& Reynolds,
  R.~T.\ 1989, \grl, 16, 1209

\bibitem[\protect\citeauthoryear{Starzl}{1931a}]{Starzl31a} Starzl,
  R.~F.\ 1931a, ``The Earthman's Burden.'' {\it Astounding Stories},
  June 1931

\bibitem[\protect\citeauthoryear{Starzl}{1931b}]{Starzl31b} Starzl,
  R.~F.\ 1931b, ``Planet of Despair.'' {\it Wonder Stories}, July 1931

\bibitem[\protect\citeauthoryear{Stern}{1992}]{Stern92} Stern,
  S.~A.\ 1992, Annual Review Astronomy \& Astrophysics, 30, 185

\bibitem[\protect\citeauthoryear{Stern et al.}{2006}]{Stern06} Stern,
  S.~A., Weaver, H.~A., Steffl, A.~J., et al.\ 2006, \nat, 439, 946

\bibitem[\protect\citeauthoryear{Stern}{2008}]{Stern08}
Stern, S. A., 2008, Space Science Reviews, 140, 3

\bibitem[\protect\citeauthoryear{Taylor}{1896}]{Taylor1896} Taylor,
  T., 1896, {\it The Mystical Hymns of Orpheus : translated from the
    Greek, and demonstrated to be the invocations which were used in
    the Eleusinian mysteries}, Bertram Dobell and Reeves and Turner,
  London

\bibitem[\protect\citeauthoryear{Tegler et al.}{2010}]{Tegler10}
  Tegler, S.~C., Cornelison, D.~M., Grundy, W.~M., et al.\ 2010, \apj,
  725, 1296

\bibitem[\protect\citeauthoryear{Tenn}{2007}]{Tenn07} Tenn, J.\ 2007,
  Journal of Astronomical History and Heritage, 10, 65

\bibitem[\protect\citeauthoryear{Tombaugh}{1946}]{Tombaugh46}
  Tombaugh, C.~W.\ 1946, Leaflet of the Astronomical Society of the
  Pacific, 5, 73

\bibitem[\protect\citeauthoryear{Tombaugh}{1960}]{Tombaugh60}
  Tombaugh, C.~W.\ 1960, \skytel, 19, 264

\bibitem[\protect\citeauthoryear{Tombaugh}{1997}]{Tombaugh97}
  Tombaugh, C.~W.\ 1997, {\it Pluto and Charon.}  University of
  Arizona Press, Tucson, eds. S. Alan Stern \& David J. Tholen, p. 15

\bibitem[\protect\citeauthoryear{van\,der\,Hucht}{2008}]{vanderHucht08}
  van der Hucht, K.~A.\ 2008, {\it Proceedings of the Twenty Sixth
    General Assembly Prague 2006}. Cambridge, UK: Cambridge University
  Press, 2008, p. 34

\bibitem[\protect\citeauthoryear{Walker \& Hardie}{1955}]{Walker55}
  Walker, M.~F., \& Hardie, R.\ 1955, Pub. Ast. Soc. Pacific, 67, 224

\bibitem[\protect\citeauthoryear{Walters}{1973}]{Walters73}
  Walters, H.\ 1973, {\it Passage to Pluto}. Nelson

\bibitem[\protect\citeauthoryear{Ward \& Canup}{2006}]{Ward06} Ward,
  W.~R., \& Canup, R.~M.\ 2006, Science, 313, 1107

\bibitem[\protect\citeauthoryear{Weaver et al.}{2005}]{Weaver05}
  Weaver, H.~A., Stern, S.~A., Mutchler, M.~J., et al.\ 2005,
  \iaucirc, 8625, 1

\bibitem[\protect\citeauthoryear{Weaver et al.}{2006}]{Weaver06}
  Weaver, H.~A., Stern, S.~A., Mutchler, M.~J., et al.\ 2006, \nat,
  439, 943

\bibitem[\protect\citeauthoryear{Weinbaum}{1935}]{Weinbaum35}
  Weinbaum, S.~G., 1935, ``The Red Peri.'' {\it Astounding Stories},
  November 1935

\bibitem[\protect\citeauthoryear{Williamson}{1933}]{Williamson33}
  Williamson, J.\ 1933, ``The Plutonian Terror.'' {\it Weird Tales},
  October 1933
\bibitem[\protect\citeauthoryear{Williamson}{1950}]{Williamson50}
  Williamson, J.\ 1933, ``{\it The Cometeers}.'' Fantasy Press

\bibitem[\protect\citeauthoryear{Yeomans}{2004}]{Yeomans04} Yeomans,
  D.~K.\ 2004, \baas, 36, 1688

\bibitem[\protect\citeauthoryear{Young et al.}{2008}]{Young08} Young,
  E.~F., French, R.~G., Young, L.~A., et al.\ 2008, \aj, 136, 1757

\end{thebibliography}
\end{document}